\documentstyle[pra,eqsecnum,aps]{revtex}

\def\displayfrac#1#2{\frac{\displaystyle #1}{\displaystyle #2}}

\begin{document}
\title{Bargmann invariants and off-diagonal geometric phases for 
multi-level quantum systems --- a unitary group approach}
\author{N. Mukunda\thanks{email: nmukunda@cts.iisc.ernet.in} \thanks{Honorary 
Professor, Jawaharlal Nehru Centre for
Advanced Scientific Research, Jakkur, Bangalore 560064}}
\address{Centre for Theoretical Studies, 
Indian Institute of Science,
Bangalore 560012, India}
\author{Arvind\thanks{email: arvind@physics.iisc.ernet.in}}
\address{ Department of Physics, Guru Nanak Dev University, Amritsar 143005, 
India}
\author{S. Chaturvedi\thanks{email: scsp@uohyd.ernet.in}}
\address{ Department of Physics, University of Hyderabad, Hyderabad 500046,
India}
\author{R. Simon\thanks{email: simon@imsc.ernet.in}}
\address{ Institute of Mathematical 
Sciences, C.I.T. Campus, Chennai 600113, 
India}
\date{\today}
\maketitle
\begin{abstract}
We investigate the geometric phases and the Bargmann invariants associated with a 
multi-level quantum systems. In particular, we show that a full set
of `gauge-invariant' objects for an $n$-level system consists of $n$
geometric phases and $\frac{1}{2}(n-1)(n-2)$ algebraically independent
$4$-vertex Bargmann invariants. In the process of establishing this result
we develop a canonical form for $U(n)$ matrices which is useful in its own
right. We show that the recently discovered `off-diagonal' geometric phases
[~N.~Manini and F.~Pistolesi, Phys. Rev. Lett. {\bf 8}, 3067 (2000)~] can be
completely analysed in terms of the basic building blocks developed
in this work. This result liberates the off-diagonal phases from the
assumption of adiabaticity used in arriving at them. 

\end{abstract}

\section{Introduction} 
The notion of geometric phase, though defined originally in the
context of adiabatic, unitary and cyclic evolution\cite{berry},
has now come to be recognised as a direct consequence of the geometry
of the complex Hilbert space and that of the associated ray space
\cite{kinematic}. The quantum kinematic\cite{kinematic} picture which
has thus emerged provides a much wider setting for the notion of the
geometric phase by rendering superfluous the various assumptions that
attended its original discovery\cite{berry}, and subsequent development
\cite{anandan,samuel,reprint}. In particular, the requirement of cyclic evolution is no longer
necessary and  it becomes possible to ascribe a geometric phase to
any open curve in the Hilbert space which has non- orthogonal unit
vectors as its end points. Further, following the quantum kinematic
approach\cite{rabei} one is led, in a natural way, to the intimate
relationship that exists between the geometric phase and the $n$-vertex
Bargmann invariants\cite{bargmann} and that between the geometric
phase and  Hamilton's theory of turns\cite{hamilton}. 
 As an application of this approach the Gouy phase
(the phase jump experienced by a focussed light beam as it crosses the focus),
 discovered  over a hundred years ago, has been shown to be a four vertex
Bargmann invariant\cite{simuk}. 

In the present work, we develop the quantum kinematic approach for the
special case of unitary evolution of an  $n$-level system. It turns out
that, in the present context, it becomes necessary to introduce Bargmann
invariants constructed out of two sets of orthonormal basis vectors. We
investigate, in detail, their properties and identify and construct
a full set of gauge-invariant building blocks for the $n$-level system.
We also develop a canonical representation of $U(n)$ matrices and bring out
its relation to the Bargmann invariants. This representation has recently
been shown to be extremely useful in parametrizing the
Cabibbo-Kobayashi-Maskawa matrix which arises in the context of
CP-violation in particle physics\cite{ckm}.

As noted above, the geometric phase becomes undefined for those open
curves in the Hilbert space which have orthogonal vectors at their
ends. A recent work by Manini and Pistolesi\cite{offdiag} addresses
itself precisely to such exceptional cases. Employing the original Berry
setting of adiabatic evolution for an $n$-level quantum system, they
introduce the concept of off-diagonal geometric phases which can be
meaningfully defined in such cases, and showed that these off-diagonal geometric phases
play an essential role in the interpretation of the findings of a recent
experiment\cite{Mano}. 

We study the off-diagonal phases of
Manini and Pistolesi within the framework of the quantum kinematic approach
and show that, in actual fact, this approach, suitably augmented, is robust
enough to accommodate the off-diagonal phases as well.
 A brief outline of this work is as follows. In section II we quickly
recapitulate the basic ingredients of the quantum kinematic approach
to  geometric phases in a general setting. In section III we specialise
the discussion to $n$-level systems and construct the gauge-invariant 
objects for this case. These turn out to be  $n$ geometric phases
and a collection of $4$-vertex Bargmann invariants. Next we consider the
problem of counting, leading to identification of a set of independent Bargmann
invariants. This necessitates a detailed analysis of the structure of the
$U(n)$ matrix group. In section IV, we develop a canonical representation
for $U(n)$ matrices which is then used, in section V, for isolating the
independent 4-vertex Bargmann invariants. In section IV, we apply the machinery developed
in the previous sections to the off-diagonal geometric phases and show how
they can be expressed in terms of the ordinary geometric phases and
the 4-vertex Bargmann invariants, thus liberating these phases from the
assumptions of adibaticity.  Section VII contains our conclusions.

\section{Geometric phases and Bargmann invariants}
We review very briefly the background and ingredients that 
go into the definition of the quantum geometric phase 
from the kinematic viewpoint, and then the Bargmann 
invariants and their properties in the generic case.
Let ${\cal H}$ be the Hilbert space of states of some
quantum system, and let ${\cal B}$ be the set of unit 
vectors in ${\cal H}$:
     \begin{eqnarray}
     {\cal B} = \{\psi\in {\cal H}\; 
     \big |\parallel\psi\parallel^2 =
     (\psi,\psi)=1\} \subset {\cal H}.
     \end{eqnarray}

\noindent
The corresponding ray space (consisting of equivalence 
classes of unit vectors which differ from one 
another by phases) and projection map are written as 
${\cal R}$ and $\pi$ respectively:
     \begin{eqnarray}
     \pi :\; {\cal B}\rightarrow {\cal R} =\mbox{space of unit rays}.
     \end{eqnarray}

\noindent
To arrive at the concept of geometric phase  we begin 
with parametrised smooth (for our purposes continuous 
and once piecewise differentiable) curves ${\cal C}$ 
of unit vectors, which may be pictured as strings 
lying in ${\cal B}$:
     \begin{eqnarray}
     {\cal C} =\{\psi(s)\in{\cal B}~|~ s_1
      \leq s\leq s_2\}\subset {\cal B}.
     \end{eqnarray}

\noindent
A gauge transformation is a smooth change of phase in 
a parameter dependent manner at each point of such a 
curve ${\cal C}$ to lead to another ${\cal C}^{\prime}$:
     \begin{eqnarray}
     {\cal C}^{\prime} = \left\{\psi^{\prime}(s) = 
     e^{i\alpha(s)}\psi(s)\; \big|\;\psi(s)\in
     {\cal C},\;\;s_1\leq s\leq s_2\right\}
     \subset {\cal B}.
      \end{eqnarray} 

\noindent
Then ${\cal C}^{\prime}$ and ${\cal C}$ share a common 
parametrised image curve $C$ in ray space:
     \begin{eqnarray}
     \pi[{\cal C}^{\prime}] =\pi [{\cal C}] = C\subset {\cal R}.
     \end{eqnarray}

\noindent
In general we permit ${\cal C}$ and even $C$ to be open curves.

The total, dynamical and geometric phases are then defined as 
follows as functionals of appropriate arguments:
     \begin{eqnarray}
     \varphi_{\mbox{tot}}[{\cal C}] &=& \arg \left(\,\psi(s_1), 
     \psi(s_2)\,\right),\nonumber\\
     \varphi_{\mbox{dyn}}[{\cal C}] &=& \mbox{Im}
     \int\limits^{s_{2}}_{s_{1}} \!ds \;\left(\psi(s),
     \frac{d\psi(s)}{ds}\right) ,\nonumber\\
     \varphi_g[C] &=& \varphi_{\mbox{tot}} \protect[{\cal C}\protect] -
     \varphi_{\mbox{dyn}} \protect[{\cal C}\protect] .
     \end{eqnarray}

\noindent
While the first two phases are functionals of ${\cal C}$ 
and do change under a gauge transformation, the geometric 
phase $\varphi_g[C]$  is gauge-invariant, which explains 
why it is written as a functional of the ray space curve
$C$.  All three phases are, however, {\em individually} 
reparametrisation invariant.

Now we turn to the Bargmann invariants and their
3connection to geometric phases.  Given any sequence of 
$n$ vectors $\psi_1,\psi_2,\ldots, \psi_n$ in ${\cal B}$,
the corresponding $n$-vertex Bargmann invariant is
     \begin{eqnarray}
     \Delta_n (\psi_1,\psi_2,\ldots, \psi_n) \equiv
           (\psi_1,\psi_2)(\psi_2,\psi_3)\;\ldots\;
     (\psi_{n-1},\psi_n)(\psi_n,\psi_1).
     \end{eqnarray}

\noindent
Here we assume in the generic case that no two successive 
vectors in the sequence are mutually orthogonal.  It is 
clear that this expression is invariant under cyclic 
permutations of the $\psi$'s, and also under independent 
phase changes of the individual vectors.  Therefore it 
is actually a quantity defined at the ray space level.  
It turns out that the phase of $\Delta_n(\psi_1,\psi_2,
\ldots, \psi_n)$ is the geometric phase for suitably 
constructed closed ray space curves obtained by joining 
each $\psi_j$ to the next $\psi_{j+1}$ (and finally
$\psi_n$ to $\psi_1$) by any so-called `null-phase curve'\cite{rabei}.  
A null-phase curve is a continuous ray-space curve such that 
for any finite connected portion of it the geometric phase 
vanishes. That is `being in phase' in the Pancharatnam sense\cite{panch}
becomes an equivalence relation on such curves.
Examples of null phase curves are ray space
geodesics (with respect to the well-known 
Fubini-Study metric\cite{page,kobayashi}), but the former are a much larger set
than the latter\cite{rabei}.  It should be emphasized that whereas the
Bargmann invariant (2.7) is defined once its `vertices',
namely the projections $\pi(\psi_1), \pi(\psi_2),
\ldots, \pi(\psi_n)$ in ${\cal R}$, are given, to interpret 
its phase as a geometric phase requires that we join each
$\psi_j$ to the next $\psi_{j+1}$ in some definite manner,
namely by some null phase curve, resulting in an `$n$-sided' 
closed figure in ${\cal R}$.

It can now be seen that as far as phases are concerned,
an $n$-vertex Bargmann invariant for $n\geq 4$ can be 
reduced to a product of $\Delta_3$ factors in the generic 
case\cite{kinematic}, so we can regard the three-vertex Bargmann invariants 
as the primitive ones.  For example we have
     \begin{eqnarray}
     \Delta_4(\psi_1,\psi_2,\psi_3,\psi_4)=
     \Delta_3(\psi_1,\psi_2, \psi_3) 
     \Delta_3(\psi_1,\psi_3,\psi_4)\big/\big|(\psi_1,\psi_3)
     \big|^2 ,
     \end{eqnarray}
    
\noindent
and more generally
     \begin{eqnarray}
     \Delta_n(\psi_1,\psi_2,\ldots, \psi_n) =\Delta_3
     (\psi_1,\psi_2,\psi_3) \Delta_{n-1}(\psi_1,\psi_3,
     \psi_4,\ldots,\psi_n)\big/\big|(\psi_1,\psi_3)\big|^2 .
     \end{eqnarray}

\noindent
Thus the geometric phases of ray space `triangles', each of 
whose sides is a null phase curve, are primitive or 
irreducible phases, and all others can be built up from them 
additively.  The  purpose in mentioning this is that in the 
particular situation we shall be dealing with later the 
primitive Bargmann invariants will turn out to be
$\Delta_4$'s rather than $\Delta_3$'s, so that situation 
will not be generic in the present sense.

\section{Gauge- invariant phases for $n$-level systems}

We now turn to a study of phases associated with $n$-level 
quantum systems.  Thus we have an $n$-dimensional complex Hilbert
space ${\cal H}_n$ describing the pure states of the system.
The unit sphere in ${\cal H}_n$, and the corresponding space of
unit rays, will be denoted by ${\cal B}_n$ and ${\cal R}_n$ 
respectively.

If we imagine that a time-dependent Hamiltonian ($n\times n$ 
hermitian matrix) is given, then at each time its complete 
orthonormal set of  eigenvectors defines an orthonormal basis 
for ${\cal H}_n$.   Assuming there are no degeneracies or level crossings, the eigenvalues can be 
arranged in, say, increasing order; and at each time this basis for ${\cal H}_n$ is 
defined upto the freedom of phase changes in each basis vector.
As time progresses this basis experiences a continuous 
unitary rotation.

     In keeping with the approach of the previous Section, 
however, we will adopt a kinematic approach here as well and 
not assume any particular Hamiltonian to be given.  Thus we
imagine that for each value of a parameter $s$ in the range
$s_1\leq s\leq s_2$ we have an orthonormal basis $\psi_j(s),
j=1,2,\ldots, n$ for  ${\cal H}_n$; and as $s$ evolves this basis experiences a continuous unitary 
evolution.  Thus we have
     \begin{eqnarray}
     (\psi_j(s), \psi_k(s)) &=& \delta_{jk},~~ j,k=1,2,\ldots,
     n;\nonumber\\
     \sum\limits^n_{j=1} \psi_j(s) \psi_j(s)^{\dag} &=& {\rm Id},~~
     s_1\leq s\leq s_2.
     \end{eqnarray}

\noindent
For ease in writing, we shall denote these vectors at the end 
points $s_1$ and $s_2$ as follows:
     \begin{eqnarray}
     \psi_j(s_1) =\psi_j,~~ \psi_j(s_2) =\phi_j .
     \end{eqnarray}

\noindent
(The orthonormal vectors $\psi_j$ here are not to be confused
with the arguments of $\Delta_n$ in eqn.(2.7)).  Our aim is to 
obtain gauge-invariant expressions and phases in this context.  
We expect to be able to construct both geometric phases 
$\varphi_g[C]$ for various $C$, and Bargmann invariants.

For each value of the index $j$, as $s$ varies from $s_1$ to 
$s_2$, the vector $\psi_j(s)$ traces out a particular continuous
parametrised curve ${\cal C}_j$ in ${\cal B}_n$:
     \begin{eqnarray}
      {\cal C}_j = \{\psi_j(s)\in{\cal B}_n~|~ s_1\leq
      s\leq s_2\}\subset {\cal B}_n,\;\;j=1,2,\ldots, n.
      \end{eqnarray}

\noindent
This curve runs from $\psi_j$ to $\phi_j$.  Its image is 
$\pi[{\cal C}_j] = C_j\subset {\cal R}_n$, and we have $n$
distinct geometric phases:
     \begin{eqnarray}
     \varphi_g[C_j]&=& \varphi_{\mbox{tot}}
     [{\cal C}_j] - \varphi_{\mbox{dyn}} [{\cal C}_j] ,\nonumber\\
     \varphi_{\mbox{tot}}[{\cal C}_j] &=&
      \arg (\psi_j, \phi_j),\nonumber\\
     \varphi_{\mbox{dyn}}[{\cal C}_j]&=& \mbox{Im}
     \int\limits_{s_{1}}^{s_{2}} ds\;
     \left(\psi_j(s),\;\frac{d\psi_j(s)}{ds}\right),\;\;
     j=1,2,\ldots, n.
     \end{eqnarray}

\noindent
Each of these geometric phases is unchanged under arbitrary 
alterations in the phase of each $\psi_j(s)$ at each parameter 
value $s$.

We turn next to the construction of Bargmann invariants, the
vertices of which are taken from the $n$ initial orthonormal
vectors $\psi_1,\ldots, \psi_n$ and the $n$ final ones 
$\phi_1,\ldots, \phi_n$.  Here we encounter an interesting 
difference compared to the generic situation discussed in the 
previous Section.  Since any two distinct $\psi_j$'s (and 
similarly any two distinct $\phi_j$'s) are orthogonal, in any 
Bargmann invariant $\Delta_n(\ldots)$ an argument $\psi_j$ 
must be followed by an argument $\phi_k$, which must be 
followed by some  $\psi_{\ell}$, and so on.  Similarly if the first argument is some $\psi$, the last 
one must be some $\phi$. Thus in the present context only even order Bargmann invariants 
$\Delta_{2\ell}$ survive, a general one being
     \begin{eqnarray}
     \Delta_{2\ell}\left(\psi_{j_{1}},\phi_{k_{1}},
     \psi_{j_{2}},\phi_{k_{2}}\ldots,
     \psi_{j_{\ell}},\phi_{k_{\ell}}\right) =
     \left(\psi_{j_{1}},\phi_{k_{1}}\right)
     \left(\phi_{k_{1}},\psi_{j_{2}}\right)
      \left(\psi_{j_{2}},\phi_{k_{2}}\right)\ldots
      \left(\psi_{j_{\ell}},\phi_{k_{\ell}}\right)
      \left(\phi_{k_{\ell}},\psi_{j_{1}}\right)
      \end{eqnarray}

\noindent
We may now regard the generic case as obtaining when every 
inner product $(\psi_j,\phi_k)$ is nonzero.  In this situation 
we find that the primitive Bargmann invariants are $\Delta_4$'s
rather than $\Delta_3$'s; for instance,
     \begin{equation}
     \Delta_6\left(\psi_{j_{1}}, \phi_{k_{1}}, \psi_{j_{2}}, 
     \phi_{k_{2}}, 
     \psi_{j_{3}}, \phi_{k_{3}}\right)=
     \Delta_4\left(\psi_{j_{1}},\phi_{k_{1}}, 
     \psi_{j_{2}}, \phi_{k_{2}}\right)
     \Delta_4 \left(\psi_{j_{1}}, \phi_{k_{2}}, 
     \psi_{j_{3}}, \phi_{k_{3}}\right)
      \big/|\left(\psi_{j_{1}}, \phi_{k_{2}}\right)\big|^2 ,
     \end{equation}

\noindent
and similarly for higher order $\Delta_{2\ell}$'s.

From this discussion  it emerges that in the present context the 
basic gauge-invariant  expressions are the $n$ geometric phases $\varphi_g[C_j]$ and the various 
4-vertex Bargmann invariants $\Delta_4\left(\psi_{j_{1}},\phi_{k_{1}},\psi_{j_{2}},
\phi_{k_{2}}\right)$.  (We are
using here quantities referring to the entire parameter range 
$s_1\leq s\leq s_2$ and to its end points, and not to any
subranges).  An important problem that now remains is to select
out of all possible $\Delta_4$'s a maximal set of
independent ones as far as phases are concerned.  For this, as a 
first step, we turn to an interesting analysis of the structure
of the unitary matrix groups $U(n)$.

\section{A canonical representation for $U(n)$ matrices, counting 
of invariant phases}

Referring to eqns.(3.1,2) we have a parameter dependent 
$n\times n$ unitary matrix describing the transition from 
the initial orthonormal basis $\{\psi_j\}$ for ${\cal H}_n$
at $s=s_1$ to the moving basis $\{\psi_j(s)\}$ at a
general $s$:
     \begin{eqnarray}
     A(s) &=& (a_{jk} (s))\;\in U(n),\nonumber\\
     a_{jk}(s) &=& (\psi_j, \psi_k(s)),~\; s_1\leq 
     s\leq s_2,\nonumber\\
     a_{jk}(s_1) &=& \delta_{jk}.
     \end{eqnarray}
 
\noindent
At $s=s_2$ we write $A(s_2) = A$:
     \begin{eqnarray}
     A &=& (a_{jk}),\nonumber\\
     a_{jk} &=& (\psi_j, \phi_k) .
     \end{eqnarray}

\noindent
The four-vertex Bargmann invariants of the type appearing 
in eqn.(3.6) are expressions involving products of matrix
elements of $A$ and their complex conjugates:
     \begin{eqnarray}
     \Delta_4(\psi_j, \phi_k, \psi_{\ell}, \phi_m)&=&
     (\psi_j, \phi_k)(\phi_k,\psi_{\ell})(\psi_{\ell},
     \phi_m)(\phi_m, \psi_j)\nonumber\\
     &=& a_{jk} a_{\ell k}^* a_{\ell m} a^*_{jm} .
     \end{eqnarray}

\noindent
Our problem is to determine  how many algebraically independent $\Delta_4$'s there are in the generic 
case in so far as their phases are concerned, and to find a convenient enumeration of 
them.  This turns out to be a somewhat intricate problem.  
After some preparation in this Section, the solution will be 
developed in the next one.

In working with $n\times n$ unitary matrices it is convenient to 
keep in mind the standard basis in ${\cal H}_n$.  Then $U(n)$
is the group of unitary transformations acting on all $n$ 
dimensions.  For $m=1,2,\ldots,n-1$, we will denote by $U(m)$ 
the unitary group acting on the first $m$ dimensions in 
${\cal H}_n$, leaving dimensions $m+1, m+2,\ldots, n$ unaffected.  
Then we have the inclusion relations (the canonical subgroup chain)
     \begin{eqnarray}
     U(1)\subset U(2)\subset U(3) \ldots \subset
     U(n-1)\subset U(n).
     \end{eqnarray}

\noindent
General matrices of $U(n), U(n-1),\ldots$ will be written as 
$A_n,A_{n-1},\ldots$.  In a matrix $A_m\in U(m)$, the last $(n-m)$
rows and columns are trivial, with ones along the diagonal and
zeroes elsewhere.  (When no confusion is likely to arise, $A_m$ 
will also denote an unbordered $m\times m$ unitary matrix).

We will now show by a recursive argument that (almost all)
elements $A_n\in U(n)$ can be expressed uniquely as $n$-fold
products
     \begin{eqnarray}
     A_n=A_n(\underline{\zeta}) A_{n-1}(\underline{\eta})
     A_{n-2}(\underline{\xi})\ldots
     A_3(\underline{\beta})A_2(\underline{\alpha})
     A_1(\chi),
     \end{eqnarray}

\noindent
where $A_n(\underline{\zeta})$ is a special $U(n)$ element
determined by an $n$-component complex unit vector 
$\underline{\zeta}\in{\cal B}_n; A_{n-1} 
(\underline{\eta})$ is a special $U(n-1)$ element
determined by an $(n-1)$-component complex unit vector 
$\underline{\eta}\in{\cal B}_{n-1}$; and so on
down to $A_2(\underline{\alpha})$ which is a special
$U(2)$ element determined  by a 2-component complex unit vector
$\underline{\alpha}\in{\cal B}_2$; and $A_1(\chi)$ is
a phase factor belonging to $U(1)$.  We are led to expect
such a representation for $A_n$ by the following argument.
Any vector $\underline{\zeta}\in{\cal B}_n$ can
be carried by a suitable $U(n)$ element into the $n^{\mbox{th}}$
vector of the standard basis, $(0,0,\ldots,0,1)^T$; and the
stability group of this vector is the subgroup
$U(n-1)\subset U(n)$ acting on the first $(n-1)$ dimensions
in ${\cal H}_n$.  Thus $U(n)$ acts transitively on ${\cal B}_n$, 
and this is just the coset space $U(n)/U(n-1)$.  Each coset is 
thus uniquely labelled by some $\underline{\zeta}\in
{\cal B}_n$.  We therefore expect that a general 
$A_n\in U(n)$ is expressible as the product 
$A_n(\underline{\zeta}) A_{n-1}$ where $\underline{\zeta}$
is the last column in $A_n$ and $A_n(\underline{\zeta})$
is a suitably chosen coset representative.  Repeating this
argument $(n-1)$ times we are led to expect the representation
(4.5).  The counting of parameters is also just right.  
Remembering that $\underline{\alpha}, \underline{\beta}, \ldots,
\underline{\xi}, \underline{\eta},\underline{\zeta}$ are 
complex unit vectors of dimensions $2,3,\ldots, n-2, n-1, n$ 
and adding the $U(1)$ phase $\chi$, the number of real
independent parameters adds up to $n^2$, the dimension of
$U(n)$.

We now present the argument leading to (4.5), yielding in 
the process the determination of $A_n (\underline{\zeta})$.
Let a generic matrix $A_n = (a_{jk})\in U(n)$ be given and
let its last ($n^{\mbox{th}}$) column be $\underline{\zeta}$:
     \begin{eqnarray}
     a_{jn} = \zeta_j,~~~ j=1,2,\ldots, n.
     \end{eqnarray}

\noindent
Multiplying $A_n$ by an $A_{n-1}$ on the right leaves this 
column unchanged.  We choose $A_{n-1}$ so as to bring the 
$n^{\mbox{th}}$ row of $A_n$ to a particularly simple form
(for ease in writing we keep using $A_n$ and $a_{jk}$ for
the $U(n)$ element obtained at each successive stage of the 
argument):
     \begin{eqnarray}
     a_{nk}&=& 0,~ \;k=1,2,\ldots, n-2;\nonumber\\
     a_{n,n-1}&=& \mbox{real positive}\nonumber\\
     &=& \left(1 - | \zeta_n|^2\right)^{1/2}.
     \end{eqnarray}

\noindent
The $A_{n-1}$ used here is arbitrary upto an $A_{n-2}$ factor 
on its right.  Having simplified the $n^{\mbox{th}}$ row of
$A_n$ in this way, we can determine all the other elements in
the $(n-1)^{\mbox{th}}$ column by imposing orthogonality of 
rows $1,2,\ldots, n-1$ to row $n$:
     \begin{eqnarray}
     a_{n,n-1} a_{j,n-1} = -\zeta_{n}^*\;\zeta_j,\;\;
     j=1,2,\ldots, n-1 .
     \end{eqnarray}

\noindent
At this point the last two columns and the last row of $A_n$ 
are known in terms of $\underline{\zeta}$.

We next use the freedom in choice of $A_{n-1}$ mentioned 
above and multiply $A_n$ on the right by a suitable $A_{n-2}$ 
(unique upto an $A_{n-3}$ on \underline{its} right) to
bring the $(n-1)^{\mbox{th}}$ row of $A_n$ to a particularly 
simple form:
     \begin{eqnarray}
     a_{n-1,k} = 0,\;\; k &=&1,2,\ldots,  n-3 ;\nonumber\\
     a_{n-1,n-2} &=& \mbox{real positive}
     \end{eqnarray}

\noindent
Normalising this row gives $a_{n-1,n-2}$:
     \begin{eqnarray}
     a_{n,n-1}\;a_{n-1,n-2} = \left(1 - |\zeta_n|^2 -
     |\zeta_{n-1}|^2\right)^{1/2}.
     \end{eqnarray}

\noindent
Next we determine all the remaining elements in the 
$(n-2)^{\mbox{th}}$ column of $A_n$ by imposing
orthogonality of rows $1,2,\ldots, n-2$ to row $(n-1)$:
     \begin{eqnarray}
     a^2_{n,n-1}\; a_{n-1,n-2} \;a_{j,n-2} =
     -\zeta^*_{n-1} \zeta_j,\;\; j=1,2,\ldots, n-2 .
     \end{eqnarray}

\noindent
At this point the last three columns and last two rows of 
$A_n$ are known in terms of $\underline{\zeta}$.

This argument can be repeated all the way until we obtain a 
matrix  $A_n(\underline{\zeta})\in U(n)$, all of whose elements 
are determined by the $n^{\mbox{th}}$ column $\underline{\zeta}$,
namely it serves as a coset representative in the coset space 
$U(n)/U(n-1)$.  (In particular the last $U(1)$ element 
$A_1(\chi)$ is used to make $a_{21}$ real positive).  After 
some algebra we obtain the result that the matrix 
$A_n(\underline{\zeta})=(a_{jk}(\underline{\zeta}))$
is uniquely determined by the conditions:
     \begin{eqnarray}
     a_{jk}(\underline{\zeta}) &=& 0,\;\;j\geq k + 2;\nonumber\\
     a_{j,j-1}(\underline{\zeta}) &=& \mbox{real positive},\;\;
     j=2,3,\ldots, n;\nonumber\\
     a_{jn}(\underline{\zeta})&=& \zeta_n,\;\; j=1,2,\ldots, n.
     \end{eqnarray}

\noindent
Thus $A_n(\underline{\zeta})$ has vanishing matrix elements 
in the lower left hand triangular portion upto two steps below
the main diagonal; nonzero matrix elements appear only one 
step below the main diagonal, and beyond.  The explicit 
expressions for the nonzero matrix elements are:  
     \begin{eqnarray}
     a_{j,j-1}(\underline{\zeta}) &=& \rho_{j-1}/\rho_j,\;\;
     j=2,3,\ldots,n;\nonumber\\
     a_{j,k}(\underline{\zeta})&=& -\zeta^*_{k+1}\zeta_j\big/
     \rho_k\;\rho_{k+1},\;\; j\leq k\leq n-1;\nonumber\\
     a_{jn}(\underline{\zeta}) &=& \zeta_n,\;\;j=1,2,\ldots,n;
     \nonumber\\
     \rho_j&=&\left(|\zeta_1|^2 + |\zeta_2|^2 +\ldots +
     |\zeta_j|^2\right)^{1/2}\nonumber\\
     &=& \left(1 - |\zeta_{j+1}|^2 -
     |\zeta_{j+2}|^2 -\ldots - |\zeta_n|^2\right)^{1/2}.
     \end{eqnarray}

\noindent
Since the quantities $\rho_j$ obey
     \begin{eqnarray}
     \rho_1 = |\zeta_1| \leq \rho_2 \leq \rho_3 \leq \ldots
     \leq \rho_{n-1} \leq \rho_n = 1,
     \end{eqnarray}

\noindent
it is evident that this determination of $A_n(\underline{\zeta})$ 
goes through with no problems as long as $\zeta_1$ is nonzero, 
ie., $\rho_1 > 0$.

It may be helpful to give the expressions for 
$A_2(\underline{\alpha})\in U(2), A_3(\underline{\beta})
\in U(3)$ determined in this way, so as to see the general 
pattern:
     \begin{mathletters}
     \begin{eqnarray}
     A_2(\underline{\alpha})&=&{\left(\begin{array}{cc}
     \displaystyle{\frac{-\alpha_2^*\alpha_1}{|\alpha_1|}}&\alpha_1\\
   |\alpha_1| &\alpha_2\end{array}\right)},\;\;|
     \alpha_1|^2 +|\alpha_2|^2 = 1;\\
     A_3(\underline{\beta})&=&{\left(\begin{array}{ccc}
     \displaystyle{\frac{-\beta_2^*\beta_1}{\rho_1\rho_2}}&
     \displaystyle{\frac{-\beta_3
     ^*\beta_1}{\rho_2}}&\beta_1\\
     \rho_1/\rho_2&\displaystyle{\frac{-\beta_3^*\beta_2}{\rho_2}}
     &\beta_2\\
     0&\rho_2&\beta_3\end{array}\right)},\;\;\nonumber\\
      &&\nonumber\\
     &&|\beta_1|^2 + |\beta_2|^2 + |\beta_3|^2=1,\nonumber\\
     &&\rho_1 = |\beta_1|,\;\rho_2 = \left(1-|\beta_3|^2
     \right)^{1/2}.
     \end{eqnarray}
     \end{mathletters}

\noindent
We notice in passing that these are not elements of $SU(2)$ 
and $SU(3)$ respectively.

Going back to the proof of eqn.(4.5), we see that it can
be recursively established; and $\chi, \underline{\alpha}, 
\underline{\beta}, \ldots, \underline{\xi}, \underline{\eta},
\underline{\zeta}$ supply us with exactly $n^2$ real 
independent parameters for $A_n$.  Of these, the 
$\frac{1}{2}n(n-1)$ independent quantities $|\alpha_1|, |\beta_1|, 
|\beta_2|,\ldots, |\zeta_1|, |\zeta_2|, \ldots, 
|\zeta_{n-1}|$ are of the modulus type, and then there are 
$\frac{1}{2}n(n+1)$ independent phases.  We can display a
general element $A_n\in U(n)$, (in particular $A$ of eqns.(4.2)) 
in the selfevident forms
     \begin{eqnarray}
     A_n &=& A_n(\underline{\zeta},\underline{\eta}, 
     \underline{\xi},\ldots,\underline{\beta},
     \underline{\alpha}, \chi)\nonumber\\
     &=& A_n (\underline{\zeta}) A_{n-1},\nonumber\\
     A_{n-1} &=& A_{n-1}
     (\underline{\eta}, \underline{\xi},\ldots,
     \underline{\beta}, \underline{\alpha}, \chi) .
     \end{eqnarray}

We are now interested in the following operation: suppose we 
premultiply and post multiply $A_n$ by two independent diagonal 
elements of $U(n)$ (a `gauge transformation' of $A_n$):
     \begin{eqnarray}
     A_n\rightarrow A^{\prime}_n &=& D_n(\theta_1,\theta_2,
     \ldots, \theta_n) \;A_n\;D_n\left(\theta^{\prime}_1,   
                             \theta^{\prime}_2,\ldots, \theta^{\prime}_n\right) ,
       \nonumber\\
      D_n(\theta_1, \theta_2,\ldots, \theta_n) &=&
      \mbox{diag} \left(e^{i\theta_{1}},
      e^{i\theta_{2}},\ldots,e^{i\theta_{n}}\right).
      \end{eqnarray}

\noindent
We would like to know: how many independent invariants can 
we construct out of  $A_n$ under these transformations, how
many of them are phases, and how can they be captured through
four-vertex Bargmann invariants ?  In the case of the matrix
$A$ of eqn.(4.2) the transformation (4.17) amounts to changing
the phase of each $\psi_j$ and each $\phi_k$ independently:
     \begin{eqnarray}
     \psi^{\prime}_j &=& e^{-i\theta_{j}}\psi_j,\nonumber\\
     \phi^{\prime}_k &=& e^{i\theta^{\prime}_{k}} 
     \phi_k,\nonumber\\
     a^{\prime}_{jk} &=& e^{i\left(\theta_{j} + 
     \theta^{\prime}_{k} \right)}a_{jk},
     \end{eqnarray}

\noindent
and we seek an independent set of invariant expressions of the 
form (4.3).

First we count the expected numbers of invariants of each kind.  
The real dimension of $U(n)$ is $n^2$.  The number of independent
$\theta$'s and $\theta^{\prime}$'s in (4.17) is $(2n-1)$,
because an overall constant phase can be attributed to either 
the left or the right diagonal factor.  Therefore the number of 
real invariants in $(n-1)^2$.  In the description (4.16) of a
general $A_n\in U(n)$, it is clear that under (4.17)
every component of each of $\underline{\alpha},\underline{\beta},
\ldots, \underline{\zeta}$ just undergoes a phase change, so the 
quantities $|\alpha_1|, |\beta_1|, |\beta_2|,\ldots,
|\zeta_1|, |\zeta_2|,\ldots,|\zeta_{n-1}|$ are $\frac{1}
{2}n(n-1)$ real independent  modulus type invariants.  Thus
there must be a balance of $\frac{1}{2}(n-1)(n-2)$ real
independent phase type invariants.  This is in agreement with 
known results\cite{harapal}.

We shall describe in the next Section a recursive procedure
by which we can pick out $\frac{1}{2}(n-1)(n-2)$ 
algebraically independent four-vector Bargmann invariants
whose phases are the expected phase invariants associated
with a general $U(n)$ matrix.

\section{Determination of independent Bargmann invariants}

We describe first how, in a recursive manner, we can
isolate the expected $\frac{1}{2}(n-1)(n-2)$ independent
gauge-invariant phases for a generic $A_n\in U(n)$
using the parametrisation (4.5), and then turn to the choice 
of an equal number of independent primitive Bargmann
invariants $\Delta_4$.

We begin with eqns.(4.5,4.16),
     \begin{eqnarray}
     A_n&=&A_n(\underline{\zeta}) A_{n-1} ,\nonumber\\
     A_{n-1}&=& A_{n-1}(\underline{\eta})
     A_{n-2}(\underline{\xi})\ldots A_2(\underline{\alpha})
     A_1(\chi)\nonumber\\
     &=& A_{n-1}(\underline{\eta},\underline{\xi},\ldots,
     \underline{\alpha}, \chi) ,
     \end{eqnarray}

\noindent
apply diagonal matrices on the left and on the right as in 
eqn.(4.17), and trace the changes that occur in $\underline
{\zeta}$ and in $A_{n-1}$:
     \begin{eqnarray}
     A^{\prime}_n&=& D_n(\theta_1, \theta_2,\ldots,\theta_n)
     A_n\;D_n\left(\theta^{\prime}_1, \theta^{\prime}_2,
      \ldots,\theta^{\prime}_n\right)\nonumber\\
     &=&A_n(\underline{\zeta}^{\prime}) A_{n-1}^{\prime}.
     \end{eqnarray}

\noindent
Our aim is to compute $\underline{\zeta}^{\prime}$ and 
$ A^{\prime}_{n-1}$.  Since the $D_n$ factors are quite 
elementary this can be carried through as follows:
     \begin{eqnarray}
     A^{\prime}_n &=& D_n(\theta_1,\theta_2,\ldots,
     \theta_n) A_n(\underline{\zeta}) A_{n-1}\;
     D_n\left(\theta^{\prime}_1,\theta^{\prime}_2,
     \ldots,\theta^{\prime}_n\right)\nonumber\\
     &=&D_n(\theta_1,\theta_2,\ldots,\theta_n)
     A_n(\underline{\zeta}) A_{n-1}\;
     D_n\left(0,0,\ldots,0,\theta^{\prime}_n\right)
     D_n\left(\theta^{\prime}_1,\theta^{\prime}_2,\ldots,
     \theta^{\prime}_{n-1},0\right)\nonumber\\
     &=&D_n(\theta_1,\theta_2,\ldots,\theta_n)
     A_n(\underline{\zeta})
     D_n\left(0,\ldots,0,\theta^{\prime}_n\right)
     A_{n-1}\;D_{n-1}\left(\theta^{\prime}_1,
     \theta^{\prime}_2,\ldots,\theta^{\prime}_{n-1}\right).
     \end{eqnarray}

\noindent
The product of the first three factors simplifies:
     \begin{eqnarray}
     D_n(\theta_1,\theta_2,\ldots,\theta_n)
     &&A_n(\underline{\zeta})
     D_n\left(0,\ldots,0,\theta^{\prime}_n\right)
     \nonumber\\
     &=&D_n(\theta_1,\theta_2,\ldots,\theta_n)
     (a_{jk}(\underline{\zeta}))
      D_n\left(0,0,\ldots,\theta_n^{\prime}\right)\nonumber\\
      &=&(b_{jk} (\underline{\zeta})),\nonumber\\
      b_{jn} (\underline{\zeta})&=&
      \zeta^{\prime}_j = e^{i\left(\theta_j+\theta^{\prime}_n
      \right)}\zeta_j ,\nonumber\\
      b_{jk}(\underline{\zeta})&=&
      e^{i\theta_{j}} a_{jk}(\underline{\zeta}),\;\;
      k=1,2,\ldots,n-1 .
      \end{eqnarray}

\noindent
Here the matrix elements $a_{jk}(\underline{\zeta})$ are given 
in eqn.(4.13), and for simplicity the $\theta$ and 
$\theta^{\prime}$ dependences of $b_{jk}$ are left implicit.  
In particular, as in eqn.(4.12),
     \begin{eqnarray}
     b_{jk}(\underline{\zeta}) = 0,\; k=1,2,\ldots, 
     j-2;\;\; j=3, 4,\ldots,n,
     \end{eqnarray}

\noindent
while
     \begin{eqnarray}
     b_{j,j-1}(\underline{\zeta}) &=& e^{i\theta_{j}} 
     a_{j,j-1} (\underline{\zeta})\nonumber\\
     &=& e^{i\theta_{j}} \rho_{j-1}/\rho_j,\;\;
     j=2,3,\ldots,n .
     \end{eqnarray}

\noindent
Thus the matrix $(b_{jk}(\underline{\zeta}))$ would have 
been $\left(a_{jk}(\underline{\zeta}^{\prime})\right)$
except for the fact that the elements $b_{j,j-1}
(\underline{\zeta})$ just below the main diagonal are
not real positive but carry phases.  But this can be
easily taken care of by extracting a suitably chosen
diagonal matrix on the right:
     \begin{eqnarray}
     \left(\,b_{jk}(\underline{\zeta})\,\right) = \left(\,a_{jk}
     (\underline{\zeta}^{\prime})\,\right)
      D_n(\theta_2, \theta_3, \ldots, \theta_n,0).
      \end{eqnarray}

\noindent
The point is that, according to the statement accompanying 
eqns.(4.12), after removal of this diagonal factor what 
remains is necessarily $A_n(\underline{\zeta}^{\prime})=
\left(a_{jk}(\underline{\zeta}^{\prime})\right)$.  
Combining the above steps we get:
     \begin{eqnarray}
      A^{\prime}_n&=&A_n(\underline{\zeta}^{\prime}) 
      A^{\prime}_{n-1}\nonumber\\
      &=& (b_{jk}(\underline{\zeta})) A_{n-1}\;
      D_{n-1}\left(\theta^{\prime}_1, \theta^{\prime}_2,
      \ldots, \theta^{\prime}_{n-1}\right)\nonumber\\
      &=&A_n(\underline{\zeta}^{\prime}) 
      D_n(\theta_2,\theta_3,\ldots,\theta_n,0)A_{n-1}\;
       D_{n-1}\left(\theta^{\prime}_1, \theta^{\prime}_2,
      \ldots, \theta^{\prime}_{n-1}\right),
      \end{eqnarray}

\noindent
so the changes induced in $\underline{\zeta}$ and in $A_{n-1}$
by the gauge transformation (5.2) are:
     \begin{mathletters}
     \begin{eqnarray}
     \zeta^{\prime}_j &=& e^{i\left(\theta_{j}+\theta^{\prime}
     _n\right)} \zeta_j,\;\;j=1,2,\ldots,n;\\
     A^{\prime}_{n-1}&=& D_{n-1}(\theta_2,\theta_3,\ldots,
      \theta_n)A_{n-1}\;
     D_{n-1}\left(\theta^{\prime}_1, \theta^{\prime}_2,\ldots,
      \theta^{\prime}_{n-1}\right).  
     \end{eqnarray}
     \end{mathletters}

\noindent
We see from the structure of this result that we can tackle 
our problem recursively: The gauge transformation(5.2) at the 
$U(n)$ level translates into the change $\underline{\zeta}
\rightarrow \underline{\zeta}^{\prime}$ given by eqn.(5.9a) 
and a gauge transformation $A_{n-1}\rightarrow A^{\prime}
_{n-1}$ at the $U(n-1)$ level given by eqn.(5.9b).  Therefore
all gauge-invariant expressions that exist at the $A_{n-1}$ or
$U(n-1)$ level survive when we move from $U(n-1)$ to
$U(n)$, and in addition as the vector $\underline{\zeta}\in
{\cal B}_n$ becomes available, new invariant phases involving
$\underline{\zeta}$ can be constructed.  The number of the
latter can be immediately computed: it is the difference
between $\frac{1}{2}(n-1)(n-2)$ and $\frac{1}{2}(n-2)(n-3)$,
namely the difference between the numbers of gauge-invariant 
phases at the $U(n)$ and the $U(n-1)$ levels, and this is 
$(n-2)$.  Therefore the number of new independent phase
invariants involving $A_n(\underline{\zeta})$, ie., 
$\underline{\zeta}$, in an essential way must be $(n-2)$.
These can now be isolated or explicitly constructed as 
follows.

From eqn.(5.9) we notice that $\theta_1$ and $\theta^{\prime}_n$
appears only in the transformation law for $\underline{\zeta}$,
not for $A_{n-1}$.  Therefore we first form the $(n-1)$ 
independent combinations $\zeta^*_j\zeta_{j+1}$ to eliminate
$\theta^{\prime}_n$ completely:
     \begin{eqnarray}
     \zeta^*_j \zeta_{j+1}\longrightarrow
     e^{-i(\theta_j-\theta_{j+1})}
     \zeta^*_j \zeta_{j+1},\;\; j=1,2,\ldots, n-1 .
     \end{eqnarray}

\noindent
Here $\theta_1$ occurs only in the transformation law for 
$\zeta^*_1 \zeta_2$, being absent as we just mentioned in the 
law for $A_{n-1}$.  Next we notice that the phases
$\theta^{\prime}_1, \theta^{\prime}_2,\ldots, \theta^{\prime}
_{n-1}$, involved in $A^{\prime}_{n-1}$, are completely absent 
in the transformation law (5.10) of $\zeta^*_j \zeta_{j+1}$.
Let us therefore look at the $(n-1)^{\mbox{th}}$ column, 
say, of $A_{n-1}$, which as is evident from eqn.(5.1)
is just the $(n-1)$ component complex unit vector
$\underline{\eta}\in{\cal B}_{n-1}$:
     \begin{eqnarray}
     A_{n-1} = \left(\begin{array}{cccc}
     \ldots&\ldots&\eta_1&0\\
     \ldots&\ldots&\eta_2&0\\
     \ldots&\ldots&\eta_{n-1}&0\\
     0&\ldots&0&1\end{array}\right).
     \end{eqnarray}

\noindent
The ``earlier'' columns of $A_{n-1}$ are more complicated, 
as is clear from the structure of $A_{n-1}$ in eqn.(5.1).
From eqn.(5.9b) we can read off the transformation law 
for the $\eta$'s under the gauge transformation (5.2):
     \begin{eqnarray}
     \eta^{\prime}_j = e^{i\left(\theta_{j+1}+\theta^{\prime}
     _{n-1}\right)} \eta_j,\;\;
     j=1,2,\ldots,n-1 .
     \end{eqnarray}

\noindent
To eliminate $\theta^{\prime}_{n-1}$ we form the $(n-2)$ 
combinations $\eta_j \eta^*_{j+1}$ which transform thus:
     \begin{eqnarray}
     \eta_j\eta^*_{j+1}\rightarrow e^{-i(\theta_{j+2}-
     \theta_{j+1})} \eta_j \eta^*_{j+1},\;\;
     j=1,2,\ldots,n-2 .
     \end{eqnarray}

\noindent
Comparing eqns.(5.10) and (5.13) we immediately obtain the 
expected $(n-2)$ independent (phase-type) invariants involving
$\underline{\zeta}\in{\cal B}_n$ in an essential manner, namely
they can be taken to be the complex quantities
     \begin{eqnarray}
     \eta_j \eta^*_{j+1} \zeta^*_{j+1} \zeta_{j+2},\;\;
     j=1,2,\ldots, n-2 .
     \end{eqnarray}

\noindent
By recursion the complete set of $\frac{1}{2}(n-1)(n-2)$
independent phase type invariants that can be formed from 
a generic matrix $A_n\in U(n)$ can be written down in terms 
of the canonical parametrisation (4.5) for $A_n$, and the
list reads:
     \begin{eqnarray}
     \alpha_j \alpha^*_{j+1} \beta^*_{j+1} \beta_{j+2}&,&
    ~~~ j=1 ;\nonumber\\
     \beta_j \beta^*_{j+1}  \gamma^*_{j+1}\gamma_{j+2}&,&
    ~~~ j=1,2;\nonumber\\
     \ldots\ldots\ldots&&\nonumber\\
     \xi_j \xi^*_{j+1} \eta^*_{j+1} \eta_{j+2}&,&
    ~~~ j=1,2,\ldots, n-3;\nonumber\\
     \eta_j \eta^*_{j+1} \zeta^*_{j+1} \zeta_{j+2}&,&
    ~~~ j=1,2,\ldots,n-2.
     \end{eqnarray}

While we have here an explicit solution to our problem, the 
difficulty is that these invariants are not directly expressed 
in terms of the matrix elements of $A_n=(a_{jk})\in U(n)$.
It is true that in our parametrisation $\underline{\zeta}$
is the last, $n^{\mbox{th}}$, column of $A_n$; but the
previous, $(n-1)^{\mbox{th}}$, column involves both
$\underline{\eta}$ and $\underline{\zeta}$; the 
$(n-2)^{\mbox{th}}$ column involves $\underline{\xi},
\underline{\eta}$ and $\underline{\zeta}$, and so on.
The task that remains is to see how to translate the 
expressions (5.15), as far as their phases are concerned,
into an algebraically equivalent set of $\frac{1}{2}(n-1)
(n-2)$  expressions formed as simply as possible out of the matrix elements of $A_n$.  We turn to this 
now, bringing in the 4-vertex Bargmann invariants of $A_n$.

As indicated in eqn.(4.3), a general 4-vertex Bargmann invariant
requires the choice of some two rows, say $j$ and $\ell$ with
$j < \ell$, and some two columns, say $k$ and $m$ with $k < m$, 
and use of the four matrix elements at their intersections:
     \begin{eqnarray}
     \Delta_{j\ell k m} \equiv a_{jk} a^*_{\ell k}
     a_{\ell m} a^*_{jm}.
      \end{eqnarray}

\noindent 
But as far as phases go, we can see in a step by step manner 
that a general $\Delta_{j\ell k m}$ reduces to a product
of factors of the simpler form
     \begin{eqnarray}
     \Delta_{jk}\equiv \Delta_{j,j+1,k,k+1}
     \end{eqnarray}

\noindent
involving some two adjacent rows and some two adjacent columns.  
This is to be understood modulo real positive definite factors
coming from the squared moduli of some of the matrix elements
of $A_n$.  The two ``recursion formulae'' that help us achieve
this simplification are:
     \begin{eqnarray}
     \Delta_{j \ell k m} &=&
     \Delta_{j \ell k m-1}\;
     \Delta_{j \ell m-1 m}/
     |a_{j,m-1}a_{\ell,m-1}|^2 ,\nonumber\\
     &=& \Delta_{j \ell-1 k m} \;\Delta_{\ell-1 \ell k m}/
     |a_{\ell-1,k} a_{\ell-1,m}|^2 .
     \end{eqnarray}

\noindent
It therefore suffices to work with the $(n-1)^2$ expressions
     \begin{eqnarray}
     \Delta_{jk} = a_{jk}\;a_{j+1,k}^*\;
     a_{j+1,k+1}\;a^*_{j,k+1},\;\;
     j,k=1,2,\ldots,n-1 ,
     \end{eqnarray}

\noindent
and their phases.  Our goal now is to (at least in principle 
and in the generic situation) express (the phases of) the 
$\frac{1}{2}(n-1)(n-2)$ complex invariants (5.15) in terms
of  (the phases of) the $(n-1)^2$ complex invariants
(5.19).  (In this process any number of real positive factors 
may intervene).  We already have here an indication that the 
$(n-1)^2$ expressions (5.19) (more exactly their phases)
cannot all be independent, the number of independent ones
being only $\frac{1}{2}(n-1)(n-2)$.  It will turn out, as we
indicate below, that these may be taken to be the $\Delta_{jk}$
for $j < k  \leq n-1$.  Again the proof is recursive in nature.

Consider the $(n-1)$ invariants (5.14) that get added to
all previous ones when we make the transition $U(n-1)\rightarrow
U(n)$ and bring in the vector $\underline{\zeta}\in{\cal B}_n$.
Instead of being expressed in terms of $\underline{\zeta}$
and $\underline{\eta}\in {\cal B}_{n-1}$, we now show that
they can be equally well expressed in terms of
$\underline{\zeta}$ 
and the penultimate, ie $(n-1)^{\mbox{th}}$, column of the
complete $U(n)$ matrix $A_n$.  Let us denote this column
vector by $\underline{w}\in{\cal B}_n$; it is orthogonal 
to $\underline{\zeta}$.  As noted earlier, it is easily
determined in terms of $\underline{\zeta}$ and $\underline
{\eta}$ or, more conveniently for our purpose, 
$\underline{\eta}$ is expressible in terms of 
$\underline{w}$ and $\underline{\zeta}$.  Starting with
    \begin{eqnarray}
     A_n&=&\left(\begin{array}{cccc}
     \ldots&\ldots&w_1&\zeta_1\\
     \ldots&\ldots&w_2&\zeta_2\\
     \ldots&\ldots&w_n&\zeta_n\end{array}\right)\nonumber\\
     &&\nonumber\\
     &=&A_n(\underline{\zeta}) A_{n-1} (\underline{\eta})
     A_{n-2},
     \end{eqnarray}

\noindent
and transposing $A_n(\underline{\zeta})$ we get
     \begin{eqnarray}
     A_{n-1}(\underline{\eta})A_{n-2} = 
     A_n(\underline{\zeta})^{\dag} A_n .
     \end{eqnarray}

\noindent
Since the factor $A_{n-2}$ does not affect the last two columns
on both sides, we can use the matrix elements (4.12,13) of
$A_n(\underline{\zeta})$ to obtain:
    \begin{eqnarray}
    \eta_j &=& \sum\limits^{j+1}_{k=1,2,\ldots} \;a_{kj}
    (\underline{\zeta})^* w_k\nonumber\\
     &=&\displayfrac{\rho_j}{\rho_{j+1}}\;w_{j+1} -
     \displayfrac{\zeta_{j+1}}{\rho_j\rho_{j+1}}
     \sum\limits^j_{k=1} \;\zeta^*_j w_j,
     \; \;j=1,2,\ldots,n-1.
     \end{eqnarray}

\noindent
The gauge  transformation laws of $\underline{\zeta}$ and $\underline{\eta}$ are given in 
eqns.(5.9a,12), while that of $\underline{w}$ is seen from eqn.(5.2) to be
     \begin{eqnarray}
     w_j\longrightarrow e^{i\left(\theta_j+\theta^{\prime}
     _{n-1}\right)} w_j,\;\;j=1,2,\ldots,n .
    \end{eqnarray}

\noindent
Naturally the relations (5.22) are consistent with these
transformation laws.  The combinations of $\underline{\eta}$
and $\underline{\zeta}$ needed in (5.14) are $\eta_j
\zeta^*_{j+1}$ for $j=1,2,\ldots,n-1$.  We see from
eqns.(5.22) that they are real linear combinations
of $w_1 \zeta_1^*, w_2\zeta^*_2,\ldots, w_{n-1}
\zeta^*_{n-1},w_n\zeta^*_n$
     \begin{eqnarray}
     \eta_j\zeta^*_{j+1}=-\displayfrac{|\zeta_{j+1}|^2}
     {\rho_j\rho_{j+1}}\;\sum\limits^j_{k=1}
     \zeta_j^* w_j + \displayfrac{\rho_j}{\rho_{j+1}}
     \;w_{j+1}\zeta^*_{j+1}.
     \end{eqnarray}

\noindent
Using both the orthogonality of $\underline{w}$ and 
$\underline{\zeta}$, and the reduction process (5.18) for
$\Delta_4$'s formed out of the last two column of $A_n$,
it is now clear that the set of complex invariants (5.14) 
can be replaced by the following set of $(n-2)\;\Delta_4$'s:
     \begin{eqnarray}
     \Delta_{j,n-1} = w_{j\cdot}\zeta^*_j w^*_{j+1}       
     \zeta_{j+1},\;\;j=1,2,\ldots,n-2 .
     \end{eqnarray}

\noindent
The known algebraic independence of the set (5.14) implies a
similar independence of these $\Delta_4$'s.

To tackle the next set of $(n-3)$ invariants $\xi_j
\eta^*_{j+1}\xi^*_{j+1}\eta_{j+2}$ for $j=1,2,\ldots,n-3$
in the list (5.15), we must bring in the $(n-2)^{\mbox{th}}$ 
column of the matrix $A_n$.  Denote this by $\underline{v}
\in{\cal B}_n$ so that
     \begin{eqnarray}
     A_n=\left(\begin{array}{cccccc}
     \cdot&\cdot&\cdot&v_1&w_1&\zeta_1\\
     \cdot&\cdot&\cdot&v_2&w_2&\zeta_2\\
     \cdot&\cdot&\cdot&\vdots&\vdots&\vdots\\
     \cdot&\cdot&\cdot&v_n&w_n&\zeta_n\\
     \end{array}\right).
     \end{eqnarray}

\noindent
Analogous to eqn.(5.21) we now have
     \begin{eqnarray}
     A_{n-2} (\underline{\xi}) A_{n-3} = 
     A_{n-1}(\underline{\eta})^{\dag}
     A_n(\underline{\zeta})^{\dag} A_n ,
     \end{eqnarray}

\noindent
from where we get expressions for $\xi_j$ in terms of
$\underline{\zeta},\underline{\eta}$ and $\underline{v}$.
This is naturally more complicated than eqn.(5.22) at 
the previous stage.  In place of the real positive factors 
$\rho_j$ defined in terms of $\underline{\zeta}$ in eqn.(4.13),
we now have similarly defined factors $\sigma_j$ in terms of
$\underline{\eta}$ occurring in the elements of $A_{n-1}
(\underline{\eta})$.  The result of comparing the 
$(n-2)^{\mbox{th}}$ columns of both sides of eqn.(5.27) is:
     \begin{eqnarray}
     \xi_j&=&\displayfrac{\sigma_j \rho_{j+1}}
     {\sigma_{j+1}\rho_{j+2}} v_{j+2} -
     \displayfrac{\sigma_j}{\sigma_{j+1}\rho_{j+1}
     \rho_{j+2}} \zeta_{j+2}\sum\limits^{j+1}
     _{\ell=1} \zeta_{\ell}^* v_{\ell} -
     \displayfrac{\eta_{j+1}}{\sigma_j\sigma_{j+1}}
     \sum\limits^j_{k=1}\displayfrac{\rho_k}{\rho_{k+1}}
     \eta^*_k v_{k+1}\nonumber\\
    &+& \displayfrac{\eta_{j+1}}{\sigma_j\sigma_{j+1}}\;
     \sum\limits^j_{k=1}\;\displayfrac{\eta^*_k\zeta_{k+1}}
     {\rho_k\rho_{k+1}}\;\sum\limits^k_{\ell=1}\;
     \zeta^*_{\ell} v_{\ell},\;~~~~~~~j=1,2,\ldots,n-2 .
     \end{eqnarray}

\noindent
Here next we can use eqn.(5.22) to go from $\underline{\eta}$ to
$\underline{w}$.  Then we form the expressions 
$\xi_j \eta^*_{{j+1}}$ and step by step work our way up to 
the invariants $\xi_j \eta^*_{j+1} \xi^*_{j+1}\eta_{j+2}$.
We can then see that apart from various real factors we 
encounter $\Delta_4$'s involving $v$'s and $w$'s,
$v$'s and $\zeta$'s and $w$'s and $\zeta$'s.  Using the
reduction rules (5.18) the $v-\zeta$ combinations can be
eliminated in favour of the other two types.  It is now
clear that apart from the $\Delta_{j,n-1}$ in eqn.(5.25)
which appeared at the previous stage, the new quantities that 
come in now are $\Delta_{j,n-2}$.  But we know in advance
that at this stage only $(n-3)$ new independent invariants
are available.  As all the rows of $\Delta_n$ are on
equal footing, we conclude that the new $\Delta_4$'s to be
added now to the previous $\Delta_{j,n-1}$ may be
taken to be
     \begin{eqnarray}
     \Delta_{j,n-2} = v_j w^*_j v_{j+1}^* w_{j+1},\;\;
     j=1,2,\ldots, (n-3) .
     \end{eqnarray}

In this manner one sees recursively that the $\frac{1}{2}
(n-1)(n-2)$ independent gauge-invariant phases in a general 
matrix $A_n\in U(n)$ are the Bargmann invariants $\Delta_{jk}$
for $j < k\leq n-1$.  In any case such a choice is permitted.  
However the actual algebraic expression of a general 
$\Delta_{jk}$ in terms of this special subset may be rather
involved, so one may freely use all $\Delta_{jk}$ in
constructing interesting gauge-invariant expressions with
various properties.

The upshot of these considerations is that the naturally
available gauge-invariant phases for the continuous unitary
evolution of an $n$-level quantum system, barring 
degeneracies and level-crossings, are $n$ geometric
phases $\varphi_g[C_j]$ as defined in eqn.(3.4),
and the $(n-1)^2$ primitive four-vertex Bargmann invariants
$\Delta_{jk}$ of eqn.(5.19); of the latter, only the
$\frac{1}{2}(n-1)(n-2)\;\Delta_{jk}$'s for
$j<k\leq n-1$ are independent.  Any composite expression 
formed out of these ingredients is of course also invariant.

\section{Off-diagonal geometric phases}

It is evident from the definitions (2.6) that while the 
dynamical phase $\varphi_{\mbox{dyn}}[{\cal C}]$ is always 
numerically well-defined once the parametrised curve
${\cal C}$ is given, the total phase $\varphi_{\mbox{tot}}
[{\cal C}]$ is only defined modulo $2\pi$, and moreover
is undefined if the vectors $\psi(s_1)$ and $\psi(s_2)$
at the end points of ${\cal C}$ are orthogonal.  These
properties naturally carry over to the geometric phase
$\varphi_g[C]$: only defined modulo $2\pi$, undefined when
$\varphi_{\mbox{tot}}[{\cal C}]$ is undefined.  The
Bargmann invariants (2.7) too share these problems of
definition as far as their phases are concerned, which
explains the limitation to generic situations.

Recently a very interesting attempt to define so-called
off-diagonal geometric phases has been made to cover
just these exceptional or problematic situations\cite{offdiag}. 
Specifically the idea is to set up gauge-invariant
phases associated with the unitary evolution of an $n$-
level quantum system, which remain well-defined even 
when one of the eigenvectors of the Hamiltonian at a
final time $t_2$, say the $k^{\mbox{th}}$ one, happens
to coincide with the $j^{\mbox{th}}$ eigenvector at
the initial time $t_1$, with $j\neq k$.  In this
situation, as $\psi_j(t_1)$ and $\psi_k(t_2)$ are the
same upto a phase, both the geometric phases $\varphi_g
[C_j]$ and $\varphi_g[C_k]$ become undefined since the
inner products $(\psi_j(t_1),\psi_j(t_2))$ and 
$(\psi_k(t_1), \psi_k(t_2))$
vanish.  We shall now briefly recall the basic quantities
introduced in this new approach, and then show that the 
usual geometric phases and Bargmann invariants as 
defined earlier can completely handle the new situation.
It is just that they must be put together in such 
combinations so that the potentially undefined factors in 
each precisely cancel one another in exceptional situations.

The notation for the evolution of an $n$ level quantum 
system is as given in Section 3.  The quantities defined
in the off-diagonal geometric phases method, when expressed
in our notations, are:
     \begin{mathletters}
      \begin{eqnarray}
     I_j &=& \exp \left\{-i\varphi_{\mbox{dyn}}
     [{\cal C}_j]\right\},\;\;j=1,2,\ldots, n ;\\
     \sigma_{jk}&=& \exp\left\{i \arg (\psi_j,\phi_k) 
      -i \varphi_{\mbox{dyn}}[{\cal C}_k]\right\},\;\;
      j\neq k ;\\
     \gamma_{jk}&=& \sigma_{jk}\;\sigma_{kj},\;\; j\neq k;\\
     \gamma_j&=& \exp\left\{i\varphi_g [C_j]\right\},\;\; 
     j=1,2,\ldots, n.
     \end{eqnarray}
     \end{mathletters}

\noindent
Of these, $I_j$ and $\sigma_{jk}$ are not gauge-invariant, but
$\gamma_{jk}$ and $\gamma_j$ are gauge-invariant.  In case
for some $j\neq k$ we have $|(\psi_j,\phi_k)|=1$, it is clear
that both $\varphi_g[C_j]$ and  $\varphi_g[C_k]$ become undefined,
but the ``off-diagonal'' quantity $\gamma_{jk}$ remains well-
defined.

The two-state or two-index quantity $\gamma_{jk}$ has been
generalised to a multi-index quantity of order $\ell$ as 
follows:
     \begin{eqnarray}
     \gamma_{j_{1}j_{2}\ldots j_{\ell}} =
     \sigma_{j_{1}j_{2}}\sigma_{j_{2}j_{3}} 
     \ldots \sigma_{j_{\ell-1}j_{\ell}} 
     \sigma_{j_{\ell}j_{1}}  ,
     \end{eqnarray}

\noindent
and this again is gauge-invariant.

We can now see that all these newly introduced gauge-invariant 
off-diagonal quantities $\gamma_{jk}, \gamma_{j_{1}j_{2}\ldots
j_{\ell}}$ are actually expressible completely in terms of
the geometric phases and Bargmann invariants for the $n$-level 
system, in carefully chosen combinations:
     \begin{eqnarray}
     \gamma_{jk}&=&\exp\left\{\,i\arg \Delta_4(\psi_j,\phi_k,
     \psi_k,\phi_j) +i\varphi_g[C_j]+i\varphi_g[C_k]\,\right\},
     \nonumber\\
     \gamma_{j_{1}j_{2}\ldots j_{\ell}} &=& 
     \exp\left\{\,i \arg
     \Delta_{2\ell}\left(\phi_{j_{1}},\psi_{j_{1}},\phi_{j_{2}},
      \psi_{j_{2}},\ldots \phi_{j_{\ell}},\psi_{j_{\ell}}\right)
     +i\varphi_g[C_{j_{1}}]+i\varphi_g\left[C_{j_{2}}\right] +\;
     \ldots \;+ i\varphi_g \left[C_{j_{\ell}}\right]\,\right\}.
     \end{eqnarray}

\noindent
In the case of $\gamma_{jk}$, for example, we see that when 
$|(\psi_j, \phi_k)|=1$ and $\varphi_g[C_j],~ \varphi_g[C_k]$
become undefined because the total phases $\varphi_{\mbox{tot}}
[{\cal C}_j]$ and $\varphi_{\mbox{tot}}[{\cal C}_k]$ are
undefined, there  are compensating factors from 
$\Delta_4(\psi_j,\phi_k,\psi_k,\phi_j)$ that precisely
cancel these parts of the individual geometric phases, so
that $\gamma_{jk}$ remains unambiguous.  The mechanism is 
similar in the case of the higher order expressions
$\gamma_{j_{1}j_{2}\ldots j_{\ell}}$

It has been shown that $\gamma_{j_{1}j_{2}\ldots j_{\ell}}$ for
$\ell \geq 4$ can be reduced to the expressions with $\ell=2$
and $\ell=3$, so these are the primitive ones.  Among these,
we can limit ourselves to choices obeying $j_{1}<j_2$ when 
$\ell=2$ and $j_1=1<j_2<j_3$ when $\ell=3$, in counting 
independent quantities.  However the upshot of our analysis 
is that we can always work with just the geometric phases 
$\varphi_g[C_j]$ and the independent $\Delta_4$'s listed
in the previous Section (but for convenience employ all
the $\Delta_{jk}$ if necessary).  All gauge-invariant 
quantities  can be built up out of them, so that conceptually
the off-diagonal geomeric phases are constructed out of
previously known familiar building blocks.

\section{Concluding remarks}
We have carried out a complete analysis of the gauge-invariant 
objects for  $n$-level quantum systems. This entails
introduction of Bargmann invariants defined over two sets of orthonormal basis 
vectors, demonstration that the primitive Bargmann invariants are 4-vertex
Bargmann invariants, and finally the identification of an algebraically
independent set of 4-vertex Bargmann invariants which turn out to be
$(n-1)(n-2)/2$ in number. In the process of achieving this task we developed
a canonical form for $U(n)$ matrices in terms of a sequence of complex
unit vectors of dimensions $n, n-1,\ldots,1$ which may be useful
in other contexts as well. Indeed, this form has already found application
in parametrising the CKM matrices which arise in the context of CP-violation
in particle physics. The gauge-invariant building blocks constructed here
are shown to provide a complete quantum kinematic picture of the 
recently discovered off-diagonal phases. The usefulness of the off-diagonal phases
 is thus extended far beyond the  
restrictive framework of adiabatic evolution. This reinforces the view that 
 the  Bargmann invariants and the traditional geometric phases,  
 and suitably constructed combinations of them, suffice 
in answering all interesting questions in this domain.

\end{document}